# Mathematical modeling of anti-tumor virus therapy:
# Regimes with complete recovery within the framework of deterministic models

Artem S. Novozhilov[1], Faina S. Berezovskaya[2], Eugene V. Koonin[1], Georgy P. Karev[1,*]

[1]National Center for Biotechnology Information, National Library of Medicine, National Institutes of Health, Bethesda, MD 20894, and [2]Department of Mathematics, Howard University, 2400 Sixth Str., Washington D.C., 20059, USA.

*To whom correspondence should be addressed. Email karev@ncbi.nlm.nih.gov

**Abstract**

A complete parametric analysis of dynamic regimes of a conceptual model of anti-tumor virus therapy is presented. The role and limitations of mass-action kinetics are discussed. A functional response, which is a function of the ratio of uninfected to infected tumor cells, is proposed to describe the spread of the virus infection in the tumor. One of the main mathematical features of ratio-dependent models is that the origin is a complicated equilibrium point whose characteristics crucially determine the main properties of the model. It is shown that, in a certain area of parameter values, the trajectories of the model form a family of homoclinics to the origin (so-called elliptic sector). Biologically, this means that both infected and uninfected tumor cells can be eliminated with time, and complete recovery is possible as a result of the virus therapy within the framework of deterministic models.

# 1. Introduction

Mathematical modeling of virus-cell interaction has a long history (e.g., (Nowak and Bangham 1996; Nowak and May 2000). Grounded in the vast and diverse theoretical epidemiology field, these mathematical models serve as valuable tools to explain empirical data, predict possible outcomes of virus infection, and propose the optimal strategy of anti-virus therapy. The unquestionable success of mathematical models of certain virus-host systems, in particular, HIV infection (Ho et al. 1995; Wei et al. 1995), provides for a reasonable hope that substantial progress can be achieved in other areas of virology as well.

Here we address a specific form of virus-cell interaction, namely, interaction of the so-called oncolytic viruses with tumors. Oncolytic viruses are viruses that specifically infect and kill cancer cells but not normal cells (Kirn and McCormick 1996; McCormick 2003; Kasuya et al. 2005). Many types of oncolytic viruses have been studied as therapeutic agents including adenoviruses, herpesviruses, and reoviruses (Kasuya et al. 2005). A specific example, which has drawn a lot of attention, is ONYX-015, an attenuated adenovirus that selectively infects tumor cells with a defect in the *p53* gene (McCormick 2003). This virus has been shown to have significant antitumor activity and has proven relatively effective at reducing or eliminating tumors in clinical trials (Kirn et al. 1998). Although safety and efficacy remain substantial concerns, several other oncolytic viruses acting on different principles, including tumor-specific transcription of the viral genome, have been developed, and some of these viruses have entered or are about to enter clinical trials.

The oncolytic effect has several possible mechanisms that yield complex results (Kasuya et al. 2005). The first such mechanism involves repeated cycles of viral replication in the tumor cells leading to rupture of the cells. The second mechanism consists in low-level virus reproduction that, however, results in the production of a cytotoxic protein, which then causes cell damage. The third mechanism involves virus infection of cancer cells that induces antitumoral immunity. Cancer cells possess weak antigens for host immune sensitization. Virus infiltration causes inflammation and lymphocyte penetration into the tumor, with the virus antigens causing increased sensitivity to tumor necrosis factor-mediated killing.

Recent technological developments have made these oncolytic viruses more tumor-specific by exploiting the tumor cell environments, but it is still unclear which virus characteristics are most important for therapeutic purposes. Viruses can be altered with respect to their rate of infection, rate of replication, or the rate at which they kill cancer cells. The current strategy for oncolytic virus therapy is to develop agents with increased safety and stronger tumor reducing effects. The expectation of using targeted replicating viruses for cancer therapy is that the virus should infect a tumor cell, replicate, and then lyse the cell. Repeated cycles of infection, virus release, spread, and reinfection of tumor cells should eventually eliminate the entire tumor. Targeting is an important safety issue, as it is clearly not desirable for the virus to replicate in normal cells of the same tissue, or in any other tissue.

The interactions between the growing tumor and the replicating virus population are highly complex and nonlinear. Hence, to precisely define the conditions that are required for successful therapy by this approach, mathematical models are needed. Clinical trials showed that the result of oncolytic virus infection on tumors can range from no apparent effect, to reduction and stabilization of the tumor load (i.e., the overall size of a tumor), to elimination of the tumor (Harrison et al. 2001). However, the simplest mathematical models describing a growing tumor under oncolytic virus fail to describe all of these outcomes; in particular, these models do not allow tumor elimination (Wodarz 2001; Wodarz and Komarova 2005). Here, we present a conceptual model of tumor cells-virus interaction, which, depending on system parameter values, exhibits various behaviors including deterministic elimination of cancer cells.

Several mathematical models that describe the evolution of tumors under viral injection were recently developed. Our model builds upon the model of Wodarz (2001) but introduces several plausible modifications. Wodarz (see also Wodarz and Komarova 2005) presented a mathematical model that describes interaction between two types of tumor cells (the cells that are infected by the virus and the cells that are not infected but are susceptible to the virus so far as they have cancer phenotype) and the immune system. Here, we consider only the direct killing of tumor cell by an oncolytic virus and, accordingly, disregard the influence of immune system. The resulting model has the general form

$$\frac{dX}{dt} = f_1(X,Y)X - g(X,Y)Y,$$
$$\frac{dY}{dt} = f_2(X,Y)Y + g(X,Y)Y,$$
(1)

where $X(t)$ and $Y(t)$ are the sizes of uninfected and infected cell populations, respectively; $f_i(X,Y)$, $i = 1, 2$, are the per capita birth rates of uninfected and infected cells; and $g(X,Y)$ is a function that describes the force of infection, i.e., the number of cells newly infected by the virus released by an infected cell per time unit. Note that there is no separate equation for free virions; it is assumed that virion abundance is proportional to infected cell abundance, which can be justified if free virus dynamics is fast compared to infected cell turnover (Nowak and May 2000). The model also assumes that, upon division of infected cells, the virus is passed on to both daughter cells. Although this is the case for the viruses that integrate into the tumor cell genome, this assumption should also be appropriate for nonintegrating viruses, because active virion production should result in a very high probability that the virus is transmitted to both daughter cells. The functions used in (Wodarz 2001) are

$$f_1(X,Y) = r_1(1-(X+Y)/K) - d,$$
$$f_2(X,Y) = r_2(1-(X+Y)/K) - a,$$
$$g(X,Y) = bX,$$
(2)

where $r_1, r_2, d, a, b, K$ are nonnegative parameters. The assumptions are that the tumor grows in a logistic fashion (with possibly different rates of growth for the uninfected and infected tumor cells), and the incidence of infection is proportional to the product $XY$; the latter assumption is based on an analogy with chemical kinetics, namely, law of mass action.

The main result of the analysis of model (1)-(2) consists in defining conditions required for maximal reduction of the tumor load. It has been suggested that "because we used deterministic model, the tumor can never go completely extinct but can be reduced to very low levels"; elimination of the tumor then might occur through stochastic effects which are not part of the model per se (Wodarz 2001). In contrast, here we show that a straightforward modification of model (1)-(2) can lead to dynamical regimes that describe elimination of the tumor cells.

Other mathematical models for tumor-virus dynamics are mainly spatially explicit models, described by systems of partial differential equations (PDE) (which is an obvious and necessary extension of ODE models inasmuch as most solid tumors have distinct spatial structure); the local dynamics, however, is usually modeled by systems of ODE that bear close resemblance to a basic model of virus dynamics (Nowak and Bangham 1996). Wu et al. (2001) modeled and compared the evolution of a tumor under different initial conditions. Friedman and Tao (2003) presented a rigorous mathematical analysis of a somewhat different model. The partial differential equation for the virus spread is the main feature that distinguishes the model of Friedman and Tao (2003) model from the model of Wu et al. (2001). Recently, Wein et al. (2003) incorporated immune response into their earlier model (Wu et al. 2001). In (Wein et al. 2003), the authors used recent preclinical and clinical data to validate their model and estimate several key parameter values. They also discussed the design of oncolytic viruses. The viruses should be designed for rapid intratumoral spread and immune avoidance, in addition to tumor-selectivity and safety. In (Wu et al. 2004), the authors made some analysis using ODE system which is a simplified approximation to their PDE model and bears some similarities to the model of Wodarz). In (Tao and Guo 2005) the authors extended the model from (Wein et al. 2003), proved global existence and uniqueness of solution in this new model, studied the dynamics of this novel therapy for cancers, and explored an explicit threshold of the intensity of the immune response for controlling the tumor. Wodarz (2003) suggested a model based on his previous work to study advantages and disadvantages of replicating versus nonreplicating viruses.

A distinct aspect of all these models is the description of the process of infection (or, if free virus dynamics is explicitly modeled, the contact process) using the law of mass action, which states that the rate of change of the uninfected cell population is proportional (if no demography effects are taken into account) to the product $XY$ (where $X$ and $Y$ are as before, or $Y$ stands for virus population if the latter is included into the model).

Under mass-action kinetics yields and the assumptions of infinitesimally short duration of contact and homogeneous mixing of the cell populations, the contact rate is proportional to the product $XY$ of the respective densities. There are situations when mass action can be a good approximation; however, in many real-life situations, it is only acceptable when $X \sim Y$, giving unrealistic rates when $X \gg Y$ or $X \ll Y$. In particular, for large populations of cells, finite and often slow spread of the virus prevents it from infecting a large number of cells per infected cell per unit of time, and a more realistic approximation of the infection process is required. The assumption underlying mass action is that the contact rate is a linear function of density $N = X + Y$. At the other extreme, the contact rate might be independent of host density. Assuming that infected and uninfected hosts are randomly mixed, this would lead to transmission function of the form $bXY/(X+Y)$. This mode of transmission is often called 'frequency-dependent' transmission (McCallum et al. 2001).

The model (1)-(2) is a version of the classical predator-prey model of a biological community; the term $bXY$ describes the simplest correspondence between prey consumption and predator production similar to the law of mass action. A crucial element in models of biological communities in the form (1) is the functional response $g(X,Y)$, i.e., the number of prey consumed per predator per time unit for given quantities of prey $X$ and predators $Y$. In the Volterra model (1931) and in model (1)-(2), this function is $bX$. Another well-known model is that of Holling (Holling 1959; for application in epidemiology see Dietz 1982; Diekmann and Heesterbeek 2000) with $g(X,Y) = bX/(1+abX)$, that takes into account the saturation effect. These two kinds of possible functional responses (and many others) do not depend on predator density ($g(X,Y) = g(X)$ and, accordingly, have been named 'prey-dependent' by Arditi and Ginzburg (1989)). In many cases, it is more realistic to assume that the functional response is ratio-dependent ($g(X,Y) = g(z)$, where $z = X/Y$ [Arditi and Ginzburg 1989]). If we consider a Holling-type function $g(z) = bz/(1+z)$, then we again obtain

$$g(X,Y)Y = b\frac{X}{X+Y}Y. \tag{3}$$

In (3) meaning of $b$ is the infection rate, i.e., the mean number of infections an infected cell can cause in a unit of time. In the terminology of epidemic models, such a rate term would be said to reflect proportional mixing as opposed to homogeneous mixing (Hwang and Kuang 2003).

The ratio-dependent models set up a challenging issue regarding their dynamics near the origin due to the fact that they are undefined at $(0,0)$. Berezovskaya et al. (2001) showed that, depending on parameter values, the origin can have its own basin of attraction in the phase space, which corresponds to the deterministic extinction of both species (Jost et al. 1999; Berezovskaya et al. 2001, 2005; Hwang and Kuang 2002). In the present context, it is clear that the ratio-dependent models display original dynamic properties that have direct connection to empirical observations of possible eradication of a tumor by virus therapy (Harrison et. al. 2001).

In the present work, we show that a plausible change of the dynamical system modeling the growth of two competing populations of cells, one of which is infected by a virus and the other one is not infected can result in a remarkable change in the model dynamics. Moreover, the additional dynamical regimes, which do not emerge in the original model, might be particularly important with respect to the underlying biological problem, the oncolytic virus therapy for cancers.

## 2. The model

We introduce our model through the incorporation of ratio-dependent process of infection (3) into the model of Wodarz (2001) (system (1)-(2)). The model based on (1) and (3), which considers two types of cells growing in logistic fashion, has the following form:

$$\frac{dX}{dt} = r_1 X \left(1 - \frac{X+Y}{K}\right) - \frac{bXY}{X+Y},$$
$$\frac{dY}{dt} = r_2 Y \left(1 - \frac{X+Y}{K}\right) + \frac{bXY}{X+Y} - aY, \quad (4)$$

where $X$ is the size of the uninfected cell population; $Y$ is the size of the infected cell population; $r_1$ and $r_2$ are the maximum per capita growth rates of uninfected and infected cells correspondingly; $K$ is the carrying capacity, $b$ is the transmission rate (this parameter also includes the replication rate of the virus); and $a$ is the rate of infected cell killing by the virus (cytotoxicity). All the parameters of the model are supposed to be nonnegative. Model (4) is subject to initial conditions $X(0) = X_0 > 0$ and $Y(0) = Y_0 > 0$. We do not include a separate equation for the virus in model (4), and the initial conditions are given for uninfected and infected cells, which imply that, at the initial moment, the system already contains some cells infected by the virus; this should be taken into account when the results of analysis of (4) are compared with clinical data.

With an appropriate change of variables $(X(t), Y(t), t) \to (x(\tau), y(\tau), \tau)$, the model can be simplified to a dimensionless form with a reduced number of independent parameters. This makes the mathematical analysis easier while preserving the essential properties of the model. There exist several formally equivalent different dimensionless forms with three parameters (which is, in the present model, the smallest possible number of parameters). The choice of the form and the specific combination of the new model parameters for the transition to the dimensionless form are defined by the biological goal of the study.

Here, our goal is to analyze system (4), mainly, with respect to its dependence on the cytotoxicity of viruses and on the force of infection, so the two parameters we are particularly interesting in are $b$ and $a$, which represent the virus characteristics that, to some extent, can be controlled. We proceed to examine the qualitative behavior of model (4) as a function of parameters. The goal is to construct the phase-parameter portrait of system (4), i.e., to divide the parameter space into domains of qualitatively (topologically) different phase behaviors.

*2.1. Phase-parameter portrait of the initial model with mass-action kinetics of the infection process*

For the sake of completeness and convenience of comparison, we present a full phase-parametric portrait of system (1)-(2) with $d = 0$, noting that the original paper of Wodarz (2001) does not present such a full analysis. We let $d = 0$ in (2), to keep the number of independent parameters as small as possible, but still preserving their non-negativity. It can be shown that the system with explicit natural mortality (i.e., $d \neq 0$) can be put into form without this additional parameter, and both of the systems have topologically equivalent phase-parametric portraits.

Rescaling model (1)-(2) by letting

$$x(\tau) = X(t)/K, \quad y(\tau) = Y(t)/K, \quad \tau = r_1 t$$

leads to the system

$$\frac{dx}{d\tau} = x(1-(x+y)) - \beta xy,$$
$$\frac{dy}{d\tau} = \gamma y(1-(x+y)) + \beta xy - \delta y, \qquad (5)$$

where $\gamma = r_2/r_1$, $\beta = bK/r_1$, $\delta = a/r_1$. There exist six topologically different domains in the parametric space $(\gamma, \beta, \delta)$ (Fig. 1). The bifurcation boundaries of the domains in Fig. 1 are $\alpha_1 = \{(\delta, \gamma, \beta): \delta = \beta\}$, $\alpha_2 = \{(\delta, \gamma, \beta): \gamma = \delta\}$, and $\alpha_3 = \{(\delta, \gamma, \beta): \gamma = \delta(\beta+1)/\beta\}$.

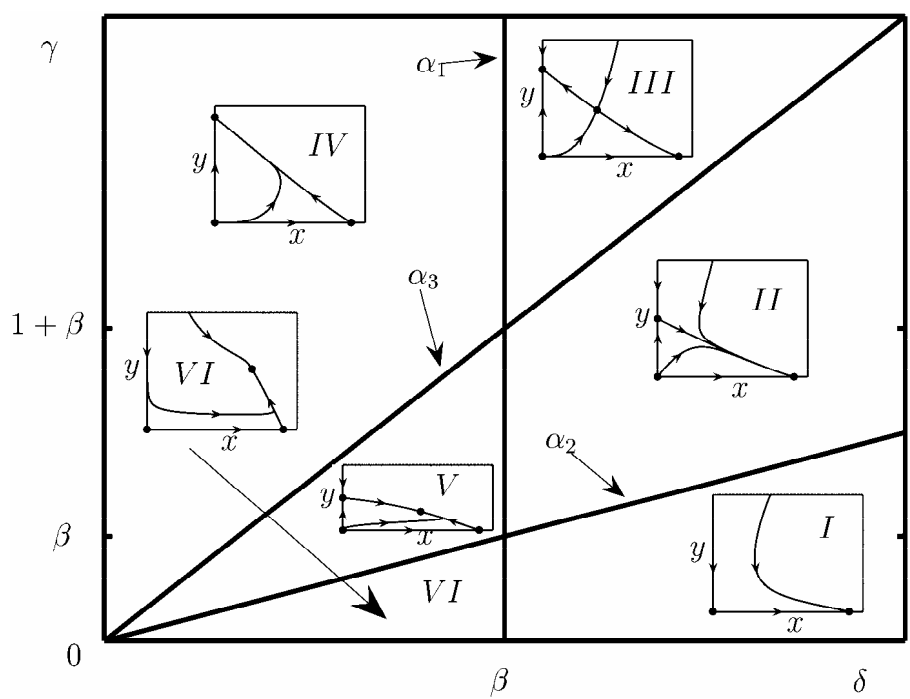

**Fig. 1.** Phase-parameter portrait of system (5) given as a cut of the positive parameter space $(\gamma, \beta, \delta)$ for an arbitrary fixed value of $\beta > 0$. The boundaries between domains correspond to changes in behavior; the corresponding equations are listed in the text. Dots represent the model equilibria.

Fig. 1 shows that there are four different regions of parameter values in which the biological interpretation differs. These are: i) domains *I* and *II* where the asymptotical state of the system is characterized by absence of infected cells; ii) domain *III* where, depending on the initial conditions, the system can find itself either in the state where all the cells are infected or in the state where all the cells are uninfected; iii) domain *IV* where the final state of the system corresponds to the absence of uninfected cells (all cells are infected); and iv) domains *V* and *VI* where there is a globally stable inner equilibrium that corresponds to coexistence of both cell populations.

We postpone the discussion of possible biological implications of the presented analysis until section 3, and at this stage, only point out that all the possible behaviors yielded by system (5) are also present in model (4) (together with additional dynamical regimes).

## 2.2. Exponential growth of the cell populations

Prior to analyzing model (4), it is worth studying its particular case when both cell populations grow unboundedly under the exponential law, i.e., formally, $1/K = 0$. The relevance of such a system is twofold. First, cancerous cells are characterized by high proliferation ability, and the exponential growth of tumors is biologically meaningful, at least, at early stages of tumorigenesis. Second, as shown below, the system with unlimited cell growth is the simplest mathematical model that possesses the property of having the elliptic sector (in biological terms, the simplest model that allows for elimination of both cell populations), and can serve as a building block to formulate and analyze more sophisticated mathematical models.

The resulting system is

$$\frac{dX}{dt} = r_1 X - \frac{bXY}{X+Y}, \quad \frac{dY}{dt} = r_2 Y + \frac{bXY}{X+Y} - aY, \tag{6}$$

where $r_1$ and $r_2$ are per capita growth rates of uninfected and infected cells, respectively (since all the parameters of the model are supposed to be nonnegative, we keep parameter $a$). Choosing another time-scale $\tau = r_1 t$, we obtain the system

$$\frac{dx}{d\tau} = x - \frac{\beta xy}{x+y}, \quad \frac{dy}{d\tau} = \gamma y + \frac{\beta xy}{x+y} - \delta y, \tag{7}$$

where $x(\tau) = X(t)$, $y(\tau) = Y(t)$, $\beta = b/r_1$, $\gamma = r_2/r_1$, and $\delta = a/r_1$. In the nondegenerate case $\gamma - \delta + \beta - 1 \neq 0$, system (7) has the only equilibrium $\mathbf{O}(0,0)$, and this equilibrium is singular.

The important mathematical peculiarity of system (7) is that the origin is a nonanalytical complicated equilibrium point. The structure of the neighborhood of point $\mathbf{O}(0,0)$ in the first quadrant of the plane $(x, y)$ and the asymptotes of trajectories for $x, y \to 0$ depend on parameter values and can change substantially with a change of parameters.

It is natural to continuously extend the determination of system (7) into the origin by changing the independent variable: $\tau \to (x+y)\tau$. Structure of the point $\mathbf{O}$ as well as the asymptotes of trajectories with $x, y \to 0$ is shown in Fig. 2 and described in Lemma 1.

**Lemma 1.** *For different positive values of parameters $\delta$, $\beta$, and $\gamma$, there exist three types of topologically different generic structures of the neighborhood of point $\mathbf{O}$ (and, accordingly, three topologically different phase portraits of system* (7)):

1) *a repelling-node sector (domain I in Fig. 2) for the parameter values $\delta < \gamma$. The phase curves of the system which tends to $\mathbf{O}$ are of the form*

$$y = Cx^{\gamma+\beta-\delta}(1+o(1)) \tag{8}$$

*if $\beta > \delta + 1 - \gamma$,*

$$y = Cx^{(\gamma-\delta)/(1-\beta)}(1+o(1)) \tag{9}$$

*if $\beta < \delta + 1 - \gamma$, where $C \neq 0$ is an arbitrary constant;*

2) *an elliptic sector* (*domain II in Fig. 2*) *composed by trajectories tending to* **O** *as* $t \to \infty$ (*with asymptotic given by* (9)), *as well as with* $t \to -\infty$ (*with asymptotic given by* (8)) *if* $\delta > \gamma$ *and* $\beta > \delta + 1 - \gamma$ (*which necessarily yields* $\beta > 1$)

3) *a saddle sector* (*domain III in Fig. 2*) *for the parameter values* $\delta > \gamma$ *and* $\beta < \delta + 1 - \gamma$;

An elliptic sector is defined as a family of homoclinics that contains no inner equilibrium (see domain *II* in Fig. 2 or Fig. A1). Using the version of the blow-up method associated with the Newton diagram (Berezovskaya 1976, 1995), Lemma 1 is proved in the Appendix. The phase-parameter portrait of (7) is given in coordinates $(\delta, \beta)$.

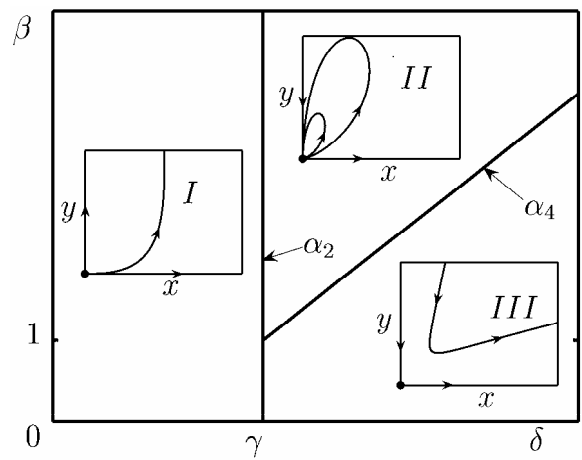

**Fig. 2.** Phase-parameter portrait of system (7) given as a cut of the positive parameter space $(\gamma, \beta, \delta)$ for an arbitrary fixed value of $\gamma > 0$. The bifurcation boundaries are $\alpha_2 = \{(\delta, \gamma) : \delta = \gamma\}$, and $\alpha_4 = \{(\delta, \gamma) : \beta = \delta + 1 - \gamma\}$

Thus, in spite of its apparent simplicity, system (7) demonstrates three different types of dynamic behavior, including the possibility to completely eliminate both cell populations in parameter domain *II* (Fig. 2). We use the results obtained for system (7) for analysis of a tumor cell-virus interaction model with the logistic growth law.

*2.3. Primary results*

If $1/K \neq 0$ in (4) and there are no infected cells ($Y_0 = 0$), then the tumor grows logistically, $X(t) \to K$ when $t \to \infty$. This assumption can be justified by graphs in (Diefenbach et al. 2001). Other mathematical forms, such as Gompertzian growth and power-law growth have been considered in other contexts, and can also be made to fit empirical data (Hart et al. 1998). In general, there is no simple universal law to describe the growth of any tumor (Retsky 2004), but we choose the logistic form since it is the simplest form whose predictions agree with the empirical data (logistic growth has been assumed also in previously analyzed models).

Re-scaling model (4) by letting

$$x(\tau) = X(t)/K, \quad y(\tau) = Y(t)/K, \quad \tau = r_1 t$$

leads to the system

$$\frac{dx}{d\tau} = x(1-(x+y)) - \frac{\beta xy}{x+y} \equiv P(x,y),$$
$$\frac{dy}{d\tau} = \gamma y(1-(x+y)) + \frac{\beta xy}{x+y} - \delta y \equiv Q(x,y), \tag{10}$$

where $\beta = b/r_1$, $\gamma = r_2/r_1$, and $\delta = a/r_1$. We proceed to study the qualitative behavior of model (10) as a function of parameters.

Let us consider the triangular region of $\mathbf{R}_2^+$

$$\Omega = \{(x,y) \in \mathbf{R}_2^+ : 0 \leq x+y \leq 1\}.$$

By examining the direction of the vector field of system (10) on the boundary of $\Omega$, it can be verified that $\Omega$ is positive invariant. Furthermore, if we assume that $x(\tau) + y(\tau) > 1$ with $x(0) + y(0) > 1$ is true for all $\tau > 0$, then

$$\frac{d(x+y)}{d\tau} = (x + \gamma y)(1-(x+y)) - \delta y \leq -\delta y. \tag{11}$$

From (11) it is followed

**Lemma 2.** *Any trajectory of system* (10) *starting within* $\mathbf{R}_2^+$ *but outside* $\Omega$ *will enter into* $\Omega$ *in a finite time.*

Hence, not only is $\Omega$ positive invariant, but it also is attractive to $\mathbf{R}_2^+$. Lemma 2 also states that any global stability in $\Omega$ is essentially the global stability in $\mathbf{R}_2^+$. We henceforth perform our mathematical analysis within the feasible domain $\Omega$.

If we choose $D(x,y) = 1/(xy)$ as a Dulac function, then

$$\frac{\partial (PD)}{\partial x} + \frac{\partial (QD)}{\partial y} = -\frac{x+\gamma y}{xy} < 0, \quad (x,y) \in \Omega$$

rules out the possibility of oscillations.

**Lemma 3.** *For any positive parameter values of* $\delta, \beta$, *and* $\gamma$, *there is no closed trajectory to system* (10).

Since closed trajectories do not exist, equilibria play a key role in determining the dynamics of the model and will be analyzed below. There are four possible equilibria $\mathbf{O}(0,0)$, $\mathbf{A}_1(1,0)$, $\mathbf{A}_2(0,(\gamma-\delta)/\gamma)$, and $\mathbf{A}_3 = (k(\beta\gamma - \delta), k(\delta - \beta))$ where $k = (\beta - 1 + \gamma - \delta)/(\beta(\gamma-1)^2)$. Equilibrium $\mathbf{O}(0,0)$ always exists. However, because neither $P(x,y)$ nor $Q(x,y)$ in (10) are analytic at this point, the linearization approach that is commonly employed to analyze the structure and the stability of this equilibrium fails. This issue received considerable attention in dynamical analysis of ecological models (e.g., see Arditi and Ginzburg 1989; Kuang and Beretta 1998). In spite of the fact that many epidemiological models with demography processes possess the same feature (e.g., Busenberg and Cook 1992; Diekmann and Heesterbeek 2000), only recently the existence and importance of this peculiarity were emphasized (Hwang and Kuang 2003; Berezovskaya et al. 2001, 2005).

The usual way to analyze this class of models involves the system

$$\frac{dx}{d\tau} = P(x, y)(x + y) \equiv P_1(x, y),$$
$$\frac{dy}{d\tau} = Q(x, y)(x + y) \equiv Q_1(x, y),$$
(12)

which is obtained from (10) with the change $d\tau \to (x + y)d\tau$. Noting that the main part of (12) coincides with system (7), we can use the results from section 2.2 to obtain the structure of a positive neighborhood of the origin of system (10). Accordingly, the possible topologically nonequivalent cases are shown in Fig. 2.

The second equilibrium $\mathbf{A}_1(1,0)$ also always exists. The local stability of $\mathbf{A}_1(1,0)$ can be examined by the regular linearization approach. The Jacobian around $\mathbf{A}_1(1,0)$ is

$$\mathbf{J}(\mathbf{A}_1) = \begin{bmatrix} -1 & -\beta - 1 \\ 0 & \beta - \delta \end{bmatrix},$$

hence the analysis of the corresponding linear system leads to proposition 1.

**Proposition 1.**
1) *If $\delta > \beta$, equilibrium $\mathbf{A}_1(1,0)$ is a stable node whereas, if $\delta < \beta$, it is a saddle;*
2) *The phase curves of the system which tend to $\mathbf{A}_1(1,0)$ are of the form*

$$y = k(x - 1)(1 + o(1)), \quad k = -(\beta - \delta + 1)/(\beta + 1).$$
(13)

*If $\mathbf{A}_1(1,0)$ is a saddle, then formula (13) and the two positive sections of the $x$-axis produced by $\mathbf{A}_1(1,0)$ determine its separatrices.*

Equilibrium $\mathbf{A}_2$ exists and belongs to $\Omega$ if $\gamma > \delta$. The Jacobian around $\mathbf{A}_2$ is

$$\mathbf{J}(\mathbf{A}_2) = \begin{bmatrix} \dfrac{\delta - \beta\gamma}{\gamma} & 0 \\ \beta - \gamma + \delta & \delta - \gamma \end{bmatrix},$$

hence the analysis of the corresponding linear system leads to proposition 2.

**Proposition 2.**
1) *If $\gamma > \delta$, equilibrium $\mathbf{A}_2$ belongs to $\Omega$. It is a saddle if $\delta > \beta\gamma$ and a stable node if $\delta < \beta\gamma$;*
2) *The phase curves of the system which tend to $\mathbf{A}_2$ are of the form*

$$y = kx(1 + o(1)) + (\gamma - \delta)/\gamma, \quad k = -\gamma(\beta - \gamma + \delta)/(\gamma(\beta - \gamma + \delta) - \delta).$$
(14)

*If $\mathbf{A}_2$ is a saddle, them formula (14) and the two positive sections of the $x$-axis produced by $\mathbf{A}_1(1,0)$ determine its separatrices.*

The fourth potential equilibrium is $\mathbf{A}_3 = (x^*, y^*)$ where

$$x^* = \frac{(\gamma\beta - \delta)(\gamma + \beta - \delta - 1)}{\beta(\gamma - 1)^2}, \quad y^* = \frac{(\delta - \beta)(\gamma + \beta - \delta - 1)}{\beta(\gamma - 1)^2}.$$

$\mathbf{A}_3$ belongs to $\Omega$ if one of the following two sets of conditions is satisfied:

$$\begin{cases} \delta > \beta \\ \gamma > \delta + 1 - \beta \\ \gamma > \delta/\beta > 1 \end{cases} \text{ or } \begin{cases} \delta < \beta \\ \gamma < \delta + 1 - \beta \\ \gamma < \delta/\beta < 1 \end{cases} \tag{15}$$

Indeed, if (15) holds:

$$x^* + y^* = 1 - \frac{\delta - \beta}{\gamma - 1} < 1.$$

The local stability of $\mathbf{A}_3$ can be examined by noting that the determinant and trace of the Jacobian around $\mathbf{A}_3$ are of the form

$$\det \mathbf{J}(\mathbf{A}_3) = \frac{(\beta - \delta)(\gamma + \beta - \delta - 1)(d\beta - \delta)}{\beta(\gamma - 1)^2}, \quad \operatorname{tr} \mathbf{J}(\mathbf{A}_3) = -\frac{(\gamma + \beta - \delta - 1)\delta}{\beta(\gamma - 1)}.$$

If the first set of conditions in (15) is satisfied, then $\det \mathbf{J}(\mathbf{A}_3) < 0$, $\operatorname{tr} \mathbf{J}(\mathbf{A}_3) < 0$, and if the second set of conditions in (15) is satisfied, then $\det \mathbf{J}(\mathbf{A}_3) > 0$, $\operatorname{tr} \mathbf{J}(\mathbf{A}_3) < 0$; hence we have the following proposition.

**Proposition 3.** *If one of the two sets of conditions (15) holds, equilibrium $\mathbf{A}_3$ belongs to $\Omega$. If $\beta > \delta$, then $\mathbf{A}_3$ is an asymptotically stable topological node, and if $\beta < \delta$, it is a saddle*

The global stability of $\mathbf{A}_3$ in case of $\beta > \delta$ follows from Proposition 3, Lemma 2, and Lemma 3.

**Proposition 4.** *The positive equilibrium $\mathbf{A}_3$ of system (10) is globally asymptotically stable in $\mathbf{R}_2^+$ if the second set of conditions (15) holds.*

*2.4. Phase-parameter portraits*

In this section, we focus on the $(x, y)$-phase and $(\delta, \beta, \gamma)$-parameter portrait of system (10). This phase-parameter portrait is obtained from the cuts on the $(\delta, \gamma)$-plane generated by fixed values of $\beta$. Four lines partition the parameter space. Their equations are listed in the caption to Fig. 3 and in Theorem 1 below. The cut of the parameter portrait on the $(\delta, \gamma)$-plane and the corresponding phase portraits critically depend on the value of $\beta$ and are different for $\beta < 1$ and $\beta > 1$. The phase-parameter portrait in $(\delta, \gamma)$-plane for the case $\beta < 1$ is almost exactly the same as for system (5) (Fig. 1). The minor difference comes from the equation for line $\alpha_3$ which, in case of system (10), is $\alpha_3 = \{(\delta, \beta, \gamma): \gamma = \delta/\beta\}$, and, consequently, the intersection point of lines $\alpha_1$ and $\alpha_3$ has coordinates $(\beta, 1)$ instead of $(\beta, 1 + \beta)$. The phase-parameter portrait of system (10) in the case $\beta > 1$ is shown in Fig. 3.

The main mathematical result of the present work is formulated in Theorem 1.

**Theorem 1.** *The space of non-negative parameters $(\delta,\beta,\gamma)$ for system (10) is subdivided into 8 domains of topologically different phase portraits belonging to $\Omega$. The cuts of the parameter space corresponding to fixed values of $\beta$ are given in Fig. 1 for $\beta<1$ and in Fig. 3 for $\beta>1$. The boundary surfaces between domains correspond to the following bifurcations for system (10):*

$\alpha_1 = \{(\delta,\beta,\gamma): \delta - \beta = 0\}$ *specifies the appearance/disappearance of equilibrium point $A_3$ and the change of the topological type of point $A_1$ (transcritical bifurcation);*

$\alpha_2 = \{(\delta,\beta,\gamma): \gamma - \delta = 0\}$ *specifies the change of the topological structure of equilibrium point $O$ with the appearance/disappearance of point $A_2$;*

$\alpha_3 = \{(\delta,\beta,\gamma): \gamma\beta - \delta = 0\}$ *for $\beta<1$ specifies the appearance/disappearance of equilibrium point $A_3$ and the change of the topological type of point $A_2$ (transcritical bifurcation).*

$\alpha_4 = \{(\delta,\beta,\gamma): \gamma - \delta - 1 + \beta = 0\}$ *for $\beta>1$ gives rise to the change of the topological structure of equilibrium point $O$ with the appearance/disappearance of point $A_3$.*

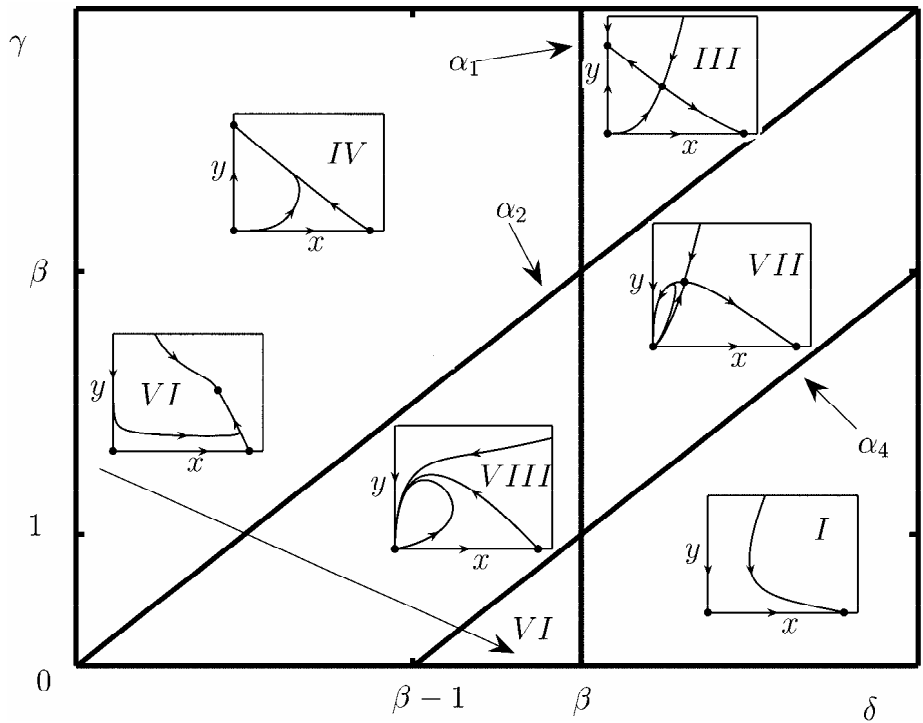

**Fig. 3.** Phase-parameter portrait of system (10) given as a cut of the positive parameter space $(\gamma,\beta,\delta)$ for an arbitrary fixed value of $\beta>1$. The dots represent the equilibria of the model. The cross-sections of the full three-dimensional parametric portrait are different for $\beta<1$ (see text and Fig. 1) and $\beta>1$ (the presented figure). The boundaries of the domains are $\alpha_1 = \{(\delta,\beta,\gamma): \delta = \beta\}$, $\alpha_2 = \{(\delta,\beta,\gamma): \gamma = \delta\}$, and $\alpha_4 = \{(\delta,\beta,\gamma): \gamma = \delta + 1 - \beta\}$.

All boundary surfaces correspond to bifurcations of co-dimension one (the total number of "connections" between parameters) in system (10) (e.g., Kuznetsov 1995). Figures 1 and 3 represent the two-dimensional cross-sections of the parameter portrait of the system for $\beta<1$ and $\beta>1$ correspondingly. Our theoretical analysis is confirmed by numerical simulations (Fig.

4 where the typical phase portraits of system (10) can be seen). The parameter values used in these simulations are listed in Table 1. Of particular interest is the occurrence of a family of homoclinics trajectories, which appear when the parameters are in domains *VII* and *VIII* in Fig. 3 (see also Fig. 4).

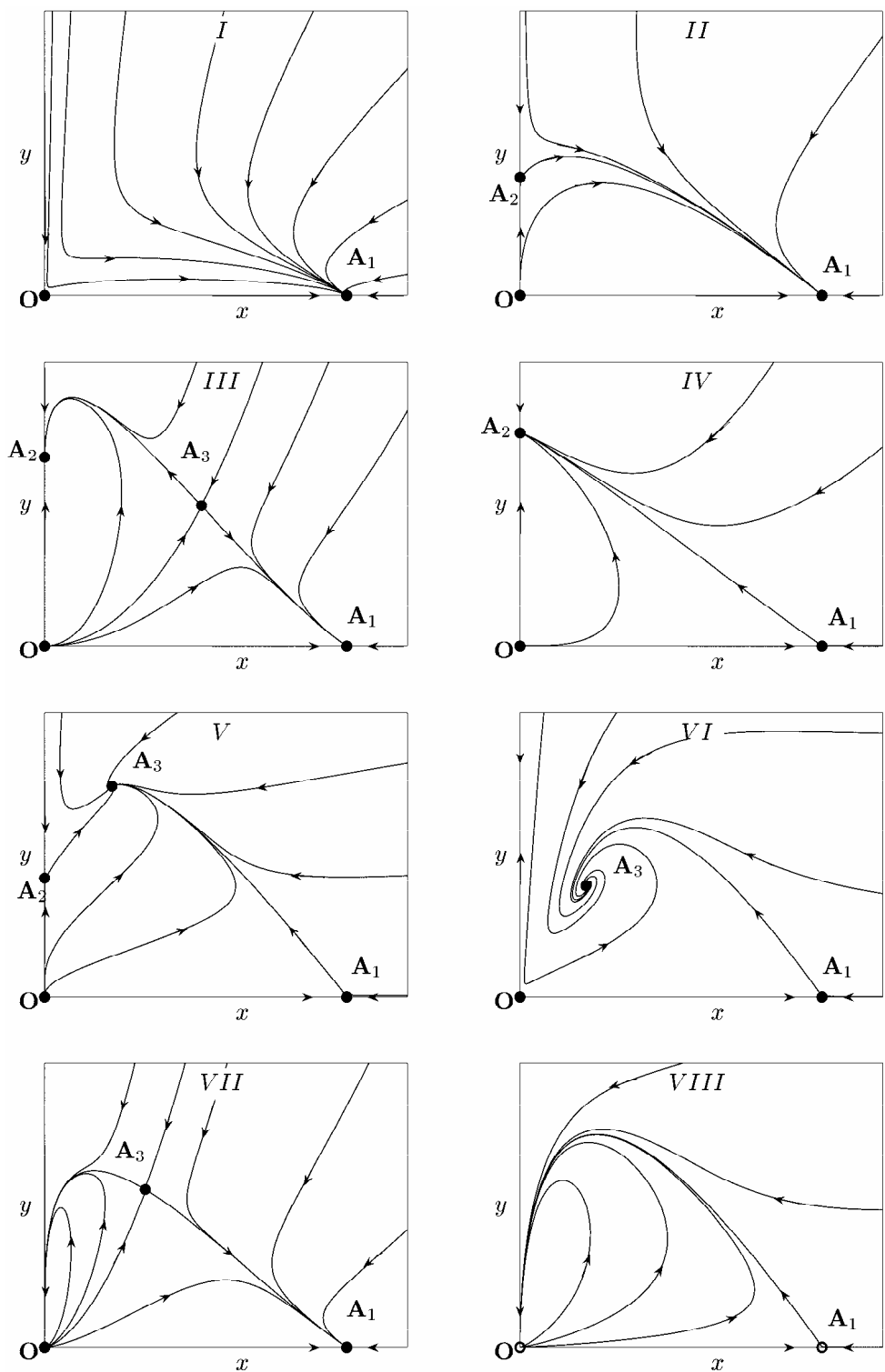

**Fig. 4.** Topologically non-equivalent phase portraits of system (10). The panels are numbered in accordance with the domains of the parameter space (Figs. 1 and 3). The parameter values used in numerical simulations are given in Table 1.

**Table 1.** Parameter values in phase portraits in Fig. 4

| Parameters: | I | II | III | IV | V | VI | VII | VIII |
|---|---|---|---|---|---|---|---|---|
| $\beta$ | 1.5 | 0.5 | 1.5 | 1.5 | 0.5 | 1.5 | 1.5 | 1.5 |
| $\delta$ | 2 | 1 | 2 | 1 | 0.3 | 1 | 2 | 1 |
| $\gamma$ | 1 | 1.8 | 2.5 | 2.5 | 0.5 | 0.3 | 1.8 | 0.7 |

Let us briefly list the order of bifurcations that appear in system (10) for typical parameter values. First, assume that $\beta > 1$ and we start in domain *I* in Fig. 3 and move counter-clockwise such that each domain is visited. In domain *I*, there are two equilibria, **O** (saddle) and $\mathbf{A}_1$ (stable node) (Fig. 4). On crossing $\alpha_4$, equilibrium $\mathbf{A}_3$ (saddle) breaks off **O**, which is accompanied by the appearance of an elliptic sector (Fig. 4, domain *VII* in Fig. 3). On line $\alpha_2$, the stable node $\mathbf{A}_2$ breaks off **O**, and the elliptic sector disappears (Fig. 4, domain *III* in Fig. 3). On line $\alpha_1$, a transcritical bifurcation occurs, with $\mathbf{A}_3$ coalescing with $\mathbf{A}_1$, and $\mathbf{A}_1$ changing its type to a saddle (Fig. 4, domain *IV* in Fig. 3). The next bifurcation occurs on $\alpha_2$ and is accompanied by the appearance of an elliptic sector; $\mathbf{A}_2$ coalesces with **O** (Fig. 4, domain *VIII* in Fig. 3). On line $\alpha_4$, the elliptic sector disappears, $\mathbf{A}_3$ (a stable equilibrium) breaks off **O** (Fig. 4, domain *VI* in Fig. 3). Finally, on $\alpha_1$, a transcritical bifurcation occurs where $\mathbf{A}_3$ coalesces with $\mathbf{A}_1$, and $\mathbf{A}_1$ changes its type to a stable node.

If $\beta < 1$ and we start from domain *I* in Fig. 1 and move counter-clockwise crossing every domain, the order of bifurcations is different. On $\alpha_2$, a saddle equilibrium $\mathbf{A}_2$ breaks off the origin, and **O** changes its type (domain *II* in Fig. 1). On $\alpha_3$, saddle equilibrium $\mathbf{A}_3$ breaks off from $\mathbf{A}_2$, and $\mathbf{A}_2$ changes its type to a stable node, which corresponds to a transcritical bifurcation (domain *III* in Fig. 1). On $\alpha_1$, $\mathbf{A}_3$ coalesces with $\mathbf{A}_1$. On $\alpha_3$, the globally stable equilibrium $\mathbf{A}_3$ breaks off $\mathbf{A}_2$, and $\mathbf{A}_2$ changes its type (domain *V* in Fig. 1). Finally, on $\alpha_2$, the saddle equilibrium $\mathbf{A}_2$ coalesces with the origin.

## 3. Interpretation of the phase-parameter portraits

Let us give an interpretation of the various behaviors of model (10) in response to changes of the initial dimensional parameters. We use the bifurcation diagram shown in Figs. 1 and 3 and, for convenience, present the bifurcation diagram for a fixed arbitrary value of $\gamma$ (Fig. 5) because the two parameters we are, mostly, interested in are *a* and *b* ($\delta$ and $\beta$ in the dimensionless form). The bifurcation diagram (Figs. 1, 3, 5) demonstrates 8 types of system dynamics depending on the values of $\delta$, $\beta$, and $\gamma$. This reflects 6 biologically distinct types of behavior because domains *I* and *II* as well as domains *V* and *VI* are indistinguishable from the biological standpoint. The results of the present analysis permit us to completely describe the parametric domains where the tumor is eliminated, the domains in which virus infection stabilizes or reduces the tumor load, and the domains in which viral therapy fails to prevent tumor growth. Using the bifurcation diagram, it is easy to predict what happens when the system crosses the boundaries of the domains.

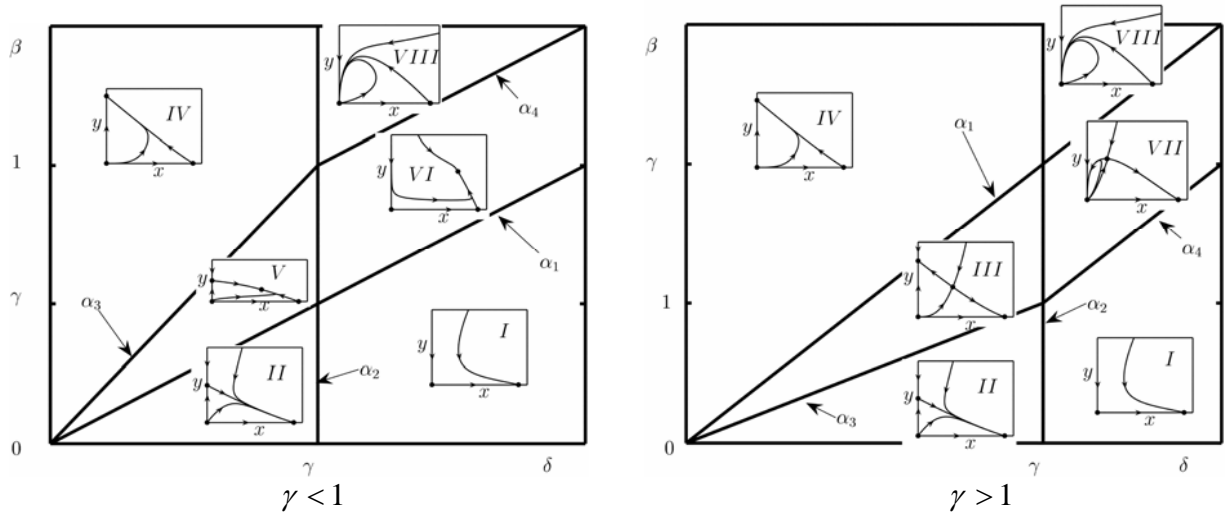

**Fig. 5.** Phase-parameter portrait of system (10) given as a cut of the positive parameter space $(\gamma, \beta, \delta)$ for an arbitrary fixed value of $0 < \gamma < 1$ and $1 < \gamma$. The boundaries of domains are listed in Theorem 1

### 3.1. Failure of viral therapy

Viral therapy fails if the eventual outcome of model dynamics is the globally asymptotically stable equilibrium $\mathbf{A}_1$ (domains *I* and *II* in Figs. 1, 3 and 5, see also Fig. 4). In terms of the dimensional parameters, the conditions for completely ineffective treatment are as follows:

$$\begin{cases} a > b > r_1 \\ r_2 - r_1 < a - b \end{cases} \quad \text{or} \quad \begin{cases} a > b \\ r_1 > b \\ r_2 / r_1 < a / b \end{cases}.$$

It has to be emphasized that the result of viral treatment in this case does not depend on the initial conditions. For any initial values of $X(0) > 0$ and $Y(0) > 0$, the trajectories of system (4) as well as system (10) tend to the same asymptotical state that would be achieved without virus administration. This suggests that, under given parameter values, even multiple, high-dose local administration of the virus to accessible tumors (a usual clinical practice) will be ineffective.

Note that it is necessary (but not sufficient) to have $a > b$, i.e., the infection rate should be less than the death rate of infected cells caused by the virus. Under these conditions, the infected cells die without having time to infect others. In the framework of the present model, this situation, in part, could be a consequence of the assumption that virus dynamics is much faster than cell dynamics, which allows us not to model viral population dynamics explicitly. If virus dynamics is not fast compared to the turnover of the cells, explicit modeling of the virus population is required.

### 3.2. The bistable situation: potential success of viral therapy

For some parameter values, we can observe a situation when the final outcome of the therapy crucially depends on the initial conditions. This is the case for domains *III* and *VII* in the bifurcation diagram (Figs. 1, 3, 4 and 5). Depending on the initial conditions, the overall tumor cell population tends either to the maximum possible tumor load ($X(t) = K$ when $t \to \infty$), or to the equilibrium $\mathbf{A}_2$ in which all cells are infected but survive (domain *III*), or to the origin, i.e., complete elimination of the tumor cells (domain *VII,* the elliptic sector). The exact conditions for the bistable situation are as follows:

$$0 < \alpha - \beta < r_2 - r_1 < \alpha - r_1$$

for the tumor elimination (domain *VII*); and

$$\begin{cases} \beta - r_1 < 0 \\ r_2/r_1 > \alpha/\beta > 1 \end{cases} \quad \text{or} \quad r_1 < \beta < \alpha < r_2$$

for stabilization of the tumor load at the equilibrium $\mathbf{A}_2$ ($Y(t) = (r_2 - a)/r_2 < K$, $X(t) = 0$ when $t \to \infty$).

Several points are worth noting with regard to the bistable situation. First, the necessary condition to have a bistable situation is $r_2 > r_1$, i.e., the maximum per capita birth rate of infected cells should exceed the maximum per capita birth rate of uninfected cells which seems to be highly unlikely unless the virus triggers cell mechanisms that favor proliferation of infected cells over uninfected cells. Second, we again have the condition $a > b$, which indicates that viruses that kill cells with high efficiency but are poorly infective would have only a limited use in anti-tumor therapy.

Formally, domains *III* and *VII* differ significantly in the possible outcomes of virus therapy because, when parameters belong to domain *III*, it is only possible to stabilize the tumor size at the value $y = (\gamma - \delta)/\gamma = (r_2 - a)/r_2$. In contrast, if the parameter values belong to domain *VII,* it is possible to eradicate the tumor (see Fig. 4), as indicated by the existence of the elliptic sector. Assuming that there is a possibility to infect tumor cells instantly, i.e., if the tumor size at the detection moment is $x$, then the initial conditions for (10) are $x(0) = x - kx$, $y(0) = kx$, where $0 < k < 1$. We can find a threshold value of $x$ such that, if the tumor size at detection is larger than $x$, the virus therapy becomes completely ineffective unless we manage to infect all tumor cells (Fig. 5). The boundary in the phase space that divides the initial conditions into dangerous (we end up in $x = 1$) and favorable (we end up either in $x = 0, y = (r_2 - a)/r_2$ or in $x = y = 0$) is the separatrice of the saddle point $\mathbf{A}_3$.

*3.3. Stabilization or reduction of tumor load*

Domains *IV, V, VI* (Figs. 1, 3, 4 and 5) are characterized by the presence of a globally stable equilibrium different from the maximal possible tumor load ($x = 1$ or $X = K$). This suggests that, by changing parameter values, we can reduce the overall tumor load to a finite minimal size. The analysis of this situation was one of the goals of Wodarz's work (2001). Note that, in this case, it is necessary to have $a < b$. The small total tumor load in the deterministic model corresponds to a real-life situation in which stochastic effects can eliminate tumor cells. Another possibility to eradicate the tumor in the deterministic setting is to lower the total tumor size to less that one cell.

Let us consider the case of $\beta < 1$ (i.e., $b < r_1$ in dimensional parameters). The overall tumor size $X(t) + Y(t)$ is given by $(r_2 - a)/r_2$ if all cells are infected (domain *IV*) and $K(1 - (b-a)/(r_1 - r_2))$ if the cell populations reach stable coexistence (domains *V* and *VI*). When $a$ increases, first, we observe decrease of the tumor size, and then, as soon as $a = a_{opt}$, the equilibrium tumor size starts to grow again (Fig. 6, the dimensionless parameter δ is used instead of the dimensional parameter *a*). The values of $a_{opt}$ correspond to the boundary between domains *IV* and *V*.

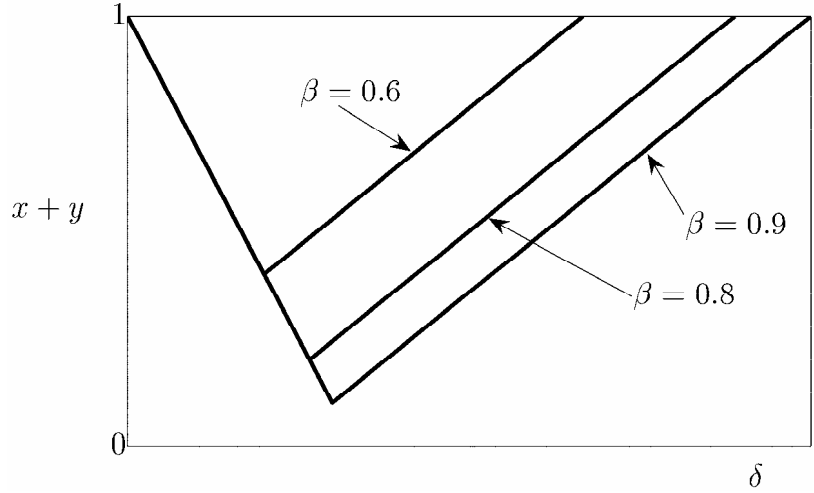

**Fig. 6.** The overall tumor load $x(\infty) + y(\infty)$ depending on viral cytotoxicity in the case of $\delta < \beta < 1$. The value of $\gamma$ is $0.3$.

The optimal virus cytotoxicity is given by

$$a_{opt} = \frac{r_2 b}{r_1},$$

and the minimal tumor size that can be reached is $K(1 - b/r_1)$. This means that, to reduce the tumor load significantly, we need to meet two conditions: $b \to r_1$ and $a \to r_2$. In dimensionless parameters, this means that we must have such values of parameters that the lines $\alpha_2$ and $\alpha_3$ coincide (Fig. 1) or at least are very close to one another. In other words, with fixed $\beta < 1$, it is impossible to choose values of $\delta$ and $\gamma$ such that the tumor size becomes arbitrarily low, and the attainable tumor size might be large enough to prevent tumor elimination due to stochastic effects, which puts into question the generality of the conclusions of Wodarz (Wodarz 2001) (see Fig. 6). Another important aspect of this situation is that, attempting to tune the parameters to maximally reduce the tumor load, we might find ourselves in highly unfavorable domains *I, II* or *III*.

*3.4. Deterministic extinction of the tumor cell population*

The most beneficial domain of parameter values in our analysis is domain *VIII* in Fig. 3. This domain corresponds to the total elimination of both cell populations (infected and uninfected) regardless of the initial conditions. This domain meets the most optimistic expectations on using replicating viruses for cancer therapy, i.e., that repeated cycles of infection, virus release, spread, and reinfection of tumor cells should eventually destroy the entire tumor. The conditions on parameter values for the system to be in this domain are

$$\begin{cases} b > a \\ b - r_1 > a - r_2 > 0 \end{cases}. \qquad (16)$$

Note that the virus has to be highly infective in comparison with its cytotoxicity.
Conditions (16) provide restrictions on parameter values to realize the regime of deterministic eradication of tumor cells; however, even if these conditions are met, on its way to extinction, the overall tumor size $X + Y$ can reach rather high values (which, with the parameters fixed, crucially depends on the initial conditions) (Fig. 7). This indicates that we must not only identify the conditions that favor tumor elimination, but also develop the optimal strategy to infect initial tumor.

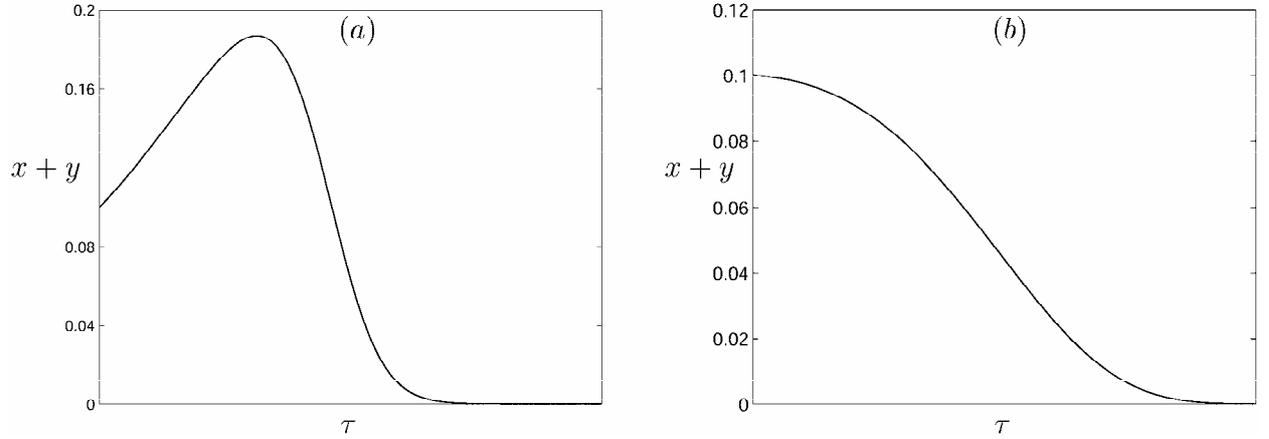

**Fig. 7.** The tumor load versus time; parameter and initial condition values are: (a) $\gamma = 0.7$, $\beta = 15$, $\delta = 10$, (b) $\gamma = 0.7$, $\beta = 1005$, $\delta = 1000$. The initial conditions are $x(0) = 0.1$, $y(0) = 0.0001$.

Let us assume that we have a possibility to instantly infect a particular part of the tumor. It is desirable to find the size of this part as a function of the model parameters such that the overall tumor size would be decreasing in the time course after infection. Since, for the maximum tumor load, $dx/d\tau + dy/d\tau = 0$, from (10) we can write down the equation for the maximum tumor size: $(x + \gamma y)(1 - (x + y)) - \delta y = 0$. Thus, if the tumor size at detection is $\tilde{x}$, we obtain the expression

$$k(\tilde{x}) = \frac{1 - \tilde{x}}{(1 - \tilde{x})(1 - \gamma) + \delta}$$

for the fraction of the cells that need to be instantly infected for the tumor load to decrease from the start of viral therapy. Since $k(\tilde{x})$ is a monotonically decreasing function ($k'(\tilde{x}) < 0$ for any $0 < \tilde{x} < 1$), we can use the value

$$k(0) = 1/(1 + \delta - \gamma) = 1/(1 + (a - r_2)/r_1) \tag{17}$$

to choose the parameters such as to to minimize the percentage of the cells that have to be infected. From (16) and (17) it is clear that, ideally, one should have $b > a \gg r_2$ and $b > a \gg r_1$. As reported in (Harrison et al. 2001), when virus-infected tumor cells are mixed with uninfected tumor cells at the time of implantation of the tumor into nude mice, 1 cell in 1000 infected with Ad337 was sufficient to prevent tumor establishment and eventually eliminate all tumor cells (Harrison et al. 2001). Our results show that this situation can be easily modeled within the framework of the model described here.

## 4. Conclusions

In this work, we present the full qualitative analysis of the deterministic model of the interaction of an oncolytic virus with tumor cells [system (10)] along with the auxiliary system

(7). System (7) is the simplest mathematical model that possesses the elliptic sector and can be used as a building block for other models.

We showed that:
i) all behaviors from Wodarz (2001) are present in (10) along with additional dynamical regimes;
ii) one of the additional dynamical regimes discovered here (domain VIII) is of particular interest from the biological and therapeutic standpoints because it demonstrates the possibility of complete eradication of the tumor by virus therapy;
iii) our model, in contrast to the model of Wodarz (2001), exhibits all possible patterns of oncolytic virus infection, i.e., no effect on the tumor, stabilization or reduction of the tumor load, and complete elimination of the tumor;
iv) the conditions on parameters were identified such that the initial infection results in immediate decrease of the tumor load and eventual elimination, and, for the given parameter values, the fraction of the tumor cells that has to be infected was found.

Although the available data are insufficient to rigorously validate the present model, it is notable that the fraction of the cells that have to be infected in order to achieve the most beneficial results within the framework of this model is comparable to the values reported in model studies on tumor implantations in the mouse model (Harrison et al. 2001). Clearly, the model described here is oversimplified, at least, in that it ignores virus population dynamics and immune system response; inclusion of parameters that characterize these and other factors may lead to more realistic models of virus-tumor interaction.

## 5. Appendix

*Proof of Lemma 1.* System (7) is analytical in all points of the plane $(x, y)$ except the origin. The positioning of its phase trajectories in the first quadrant is identical to those of the polynomial system

$$\frac{dx}{d\tau} = x(x+y) - \beta xy,$$
$$\frac{dy}{d\tau} = \gamma y(x+y) + \beta xy - \delta y(x+y)$$
(A.1)

obtained from (7) by a change of the independent variable $\tau \to (x+y)\tau$. System (A.1) has a complicated equilibrium point at the origin (because both eigenvalues are equal to zero) which is investigated below by methods developed in (Berezovskaya 1976, 1995).

The first step consists in a change of variables in system (A.1) (see also Jost et al. 1999)

$$x = x, \quad u = y/x, \quad x\,d\tau = ds$$

that transforms in a non-degenerate way the first quadrant of the $(x, y)$-plane, except $x = 0$ into the first quadrant of the $(x, u)$-plane and blows-up the point **O** into the $u$-axis. The resulting system is

$$\frac{dx}{ds} = x(1 + (1-\beta)u),$$

$$\frac{du}{ds} = u(\gamma - 1 + \beta - \delta + (\gamma - 1 + \beta - \delta)u),$$

which has only one equilibrium point on the $u$-axes: $\mathbf{O}_1(0,0)$ (we are only interested in nonnegative equilibria). The eigenvalues of this point are $\lambda_1(\mathbf{O}_1) = 1$, and $\lambda_2(\mathbf{O}_1) = \gamma - 1 + \beta - \delta$, hence this point is non-degenerate if $\gamma \neq \delta + 1 - \beta$. If $\gamma > \delta + 1 - \beta$ then $\mathbf{O}_1(0,0)$ is an unstable node, while it is a saddle if $\gamma < \delta + 1 - \beta$ (see also Fig. 2 and A1). In the case of unstable node, $\mathbf{O}_1(0,0)$ is the source of a family of trajectories with asymptotes

$$u = Cx^{\gamma + \beta - \delta - 1}(1 + o(1))$$

where $C \neq 0$ is an arbitrary constant. In coordinates $(x, y)$ this family is transformed into family (8).

We now repeat the blow-up procedure to study the behavior of the system close to the $y$-axes:

$$y = y, \quad v = x/y, \quad y\, d\tau = ds.$$

This transformation is non-degenerate for all values of $x, y$ except $y = 0$ and the point $\mathbf{O}(0,0)$ blows-up into the $v$-axes. In variables $(v, y)$ we obtain the system

$$\frac{dv}{ds} = v(1 - \beta - \gamma + \delta + (1 - \beta - \gamma + \delta)v),$$

$$\frac{dy}{ds} = y(\gamma - \delta - (\beta + \gamma - \delta)v).$$
(A.2)

This system has equilibrium $\mathbf{O}_2(0,0)$ on the $v$-axes with eigenvalues $\lambda_1(\mathbf{O}_2) = \delta + 1 - \beta - \gamma$ and $\lambda_2(\mathbf{O}_2) = \gamma - \delta$. Depending on the parameter values the following cases are realized in system (A.2) in the first quadrant of the $(v, y)$ plane (Fig. A1):

(i) $\mathbf{O}_2(0,0)$ is a saddle with $v$-axes stable manifold if $\gamma > \delta$ and $\beta > \delta + 1 - \gamma$;
(ii) $\mathbf{O}_2(0,0)$ is a unstable node if $\beta < \delta + 1 - \gamma$ and $\gamma > \delta$;
(iii) $\mathbf{O}_2(0,0)$ is a saddle with $y$-axes stable manifold if $\gamma < \delta$ and $\beta < \delta + 1 - \gamma$;
(iv) $\mathbf{O}_2(0,0)$ is a stable node if $\gamma < \delta$ and $\beta > \delta + 1 - \gamma$;

In cases (ii) and (iv) equilibrium $\mathbf{O}_2(0,0)$ is the source of the family of trajectories

$$v = Cy^{(\delta + 1 - \beta - \gamma)/(\gamma - \delta)}(1 + o(1)).$$

Returning to original coordinates we obtain family (9).

Assembling together the obtained results and returning to the initial variables $(x, y)$ as shown in Fig. A1, we obtain different topological structures of the complicated point $\mathbf{O}(0,0)$ in the first quadrant of the plane $(x, y)$ depending on the system parameters.

Note that the phase portraits in the neighborhood of $\mathbf{O}$ in Figs. A1a and A1d are topologically equivalent and differ only in the asymptotes of characteristic trajectories.

Therefore, there are only three topologically different structures in the plane $(x, y)$ in non-degenerate cases; they are presented in Fig. A1.

Proof of Lemma 1 is now complete.

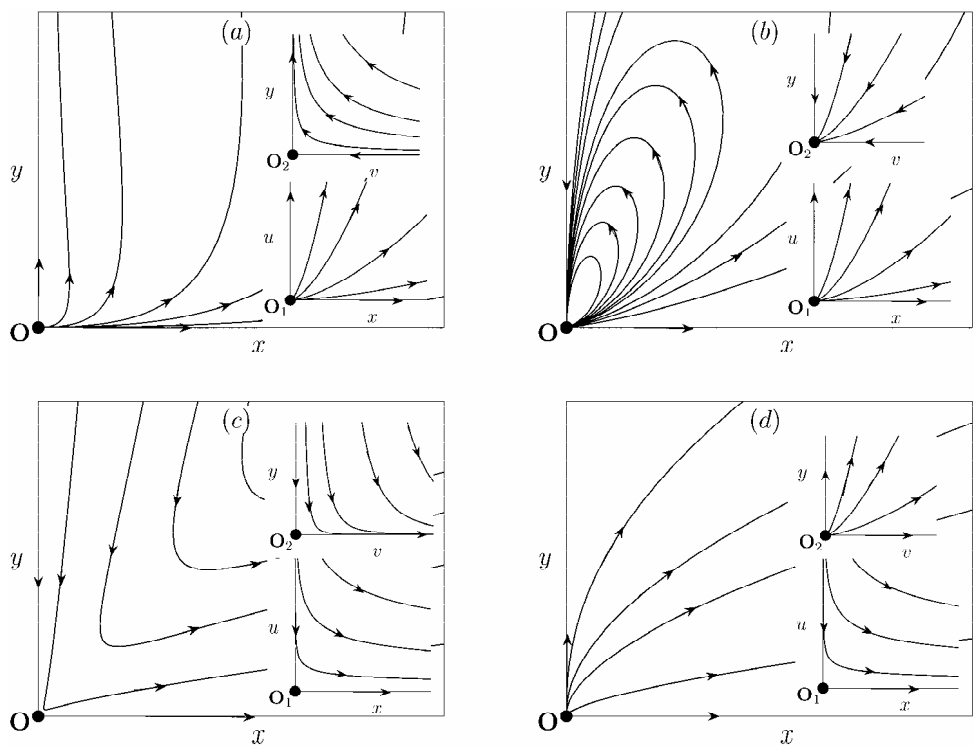

**Fig. A1.** Phase portraits of the positive neighborhood of $\mathbf{O}$, which correspond different domains in Fig. 2 together with phase portraits of auxiliary systems (see the text), invoked for analysis of singular equilibrium $\mathbf{O}$. The parameter values used in numerical simulation are given in Table A1

**Table A1.** Parameter values in phase portraits of Fig. A1

| Parameters: | (a) | (b) | (c) | (d) |
|---|---|---|---|---|
| $\beta$ | 1.5 | 1.5 | 1.5 | 0.5 |
| $\delta$ | 1 | 1 | 1 | 1 |
| $\gamma$ | 2 | 0.7 | 0.3 | 1.3 |